\documentclass[preprint,12pt]{elsarticle}

\usepackage{amssymb}
\usepackage{amsmath}
 \usepackage{multirow}
\usepackage{lineno}

\journal{Journal of High Energy Astrophysics}
\DeclareUnicodeCharacter{0301}{\'{e}} 
\begin{document}

\begin{frontmatter}



\title{Search for Magnetic Monopoles with the Complete
ANTARES Dataset}


\author[a,b]{A.~Albert}
\author[c]{S.~Alves}
\author[d]{M.~Andr\'e}
\author[e]{M.~Ardid}
\author[e]{S.~Ardid}
\author[f]{J.-J.~Aubert}
\author[g]{J.~Aublin}
\author[g]{B.~Baret}
\author[h]{S.~Basa}
\author[g]{Y.~Becherini}
% ...

\affiliation[a]{country={Université de Strasbourg, CNRS, IPHC UMR 7178, F-67000 Strasbourg, France}}
\affiliation[b]{country={Université de Haute Alsace, F-68100 Mulhouse, France}}

\affiliation[c]{country={IFIC - Instituto de Fı́sica Corpuscular (CSIC - Universitat de València) c/ Catedrático José Beltrán, 2 E-46980 Paterna, Valencia, Spain  }}

\affiliation[d]{country={Technical University of Catalonia, Laboratory of Applied Bioacoustics, Rambla Exposició, 08800 Vilanova i la Geltrú, Barcelona, Spain }}
\affiliation[e]{country={Institut d’Investigació per a la Gestió Integrada de les Zones Costaneres (IGIC) - Universitat Politècnica de València. C/ Paranimf 1, 46730 Gandia, Spain }}

\affiliation[f]{country={Aix Marseille Univ, CNRS/IN2P3, CPPM, Marseille, France }}
\affiliation[g]{country={Université Paris Cité, CNRS, Astroparticule et Cosmologie, F-75013 Paris, France }}
\affiliation[h]{country={Aix Marseille Univ, CNRS, CNES, LAM, Marseille, France}}
% ...

\cortext[cor1]{ahmed\_eddymaoui@um5.ac.ma}
\fntext[label2]{Deceased}




\begin{abstract}
This study presents a novel search for magnetic monopoles using data collected over a 14 year period (2008-2022) by the ANTARES neutrino telescope. The interaction of magnetic monopoles with matter was modeled according to Kazama, Yang, and Goldhaber cross-section. Upper limits on the flux of magnetic monopoles are obtained for velocities both above and below the Cherenkov threshold.  No events consistent with the passage of magnetic monopoles were detected, 
enabling the setting of an upper flux limit for relativistic magnetic monopoles of the order of \(  10^{-18} \, \text{cm}^{-2} \, \text{s}^{-1} \, \text{sr}^{-1} \).
\end{abstract}





\end{frontmatter}



\tableofcontents

\section{Introduction}

Magnetic charges and currents have been introduced to reinstate symmetry within Maxwell’s equations for both magnetic and electric fields. Paul Dirac demonstrated in 1931 \cite{Dirac:1931kp} that the incorporation of magnetic monopoles offers a profound explanation for electric charge quantization. This theoretical landscape was expanded by Grand Unified Theories (GUTs), which suggested that magnetic monopoles were created during the symmetry-breaking phase transition that followed the Big Bang. Furthermore, GUT magnetic monopoles may have a mass greater than 10$^{14}$ GeV \cite{PhysRevLett.58.1707}. Notably, magnetic monopoles emit significantly more light than muons in transparent mediums, such as water and ice, making their signals distinguishable from atmospheric muons in a Cherenkov telescope, such as ANTARES. 
\noindent The analysis of this paper utilizes ANTARES data collected from 2008 to 2022, covering a total of 3286 days of livetime. The earlier findings \cite{Adri_n_Mart_nez_2012}\cite{Albert_2017}\cite{20221} are enhanced by this analysis. 
 The ANTARES neutrino telescope \cite{Ageron_2011} was completed in 2008 and fully decommissioned, with retrieval from the sea completed in the summer of 2022. The data presented in this analysis constitutes the final contribution to the search for magnetic monopoles.\\
\noindent This paper provides a brief overview of the ANTARES neutrino telescope in section 2. Section 3 introduces the physics of magnetic monopoles and the light yield produced by their passage through the detector. Section 4 details the simulation framework, trigger algorithms, and reconstruction techniques, while section 5 focuses on the event selection criteria. Section 6 presents the statistical methods. Finally, the results obtained are presented and discussed in section 7.

\section{The ANTARES Detector}
ANTARES was a large neutrino telescope located in the Mediterranean Sea, about 40 km off the coast of Toulon, France \cite{Ageron_2011}. The detector was composed of a three-dimensional array of about 900 optical modules (OMs) deployed along 12 vertical detection lines anchored to the seafloor at a depth of approximately 2.5 km.
Each line comprised 25 floors, each equipped with three OMs, which were distributed along the line to detect the Cherenkov light induced by the passage of charged particles produced in neutrino interactions with seawater. The telescope covered an instrumented volume of about 0.02 cubic $\text{km}^3$. ANTARES was primarily designed to detect high-energy neutrinos from astrophysical sources \cite{Albert_2021_1}. These neutrinos are produced in extreme cosmic environments, such as microquasars and supernova remnants.
In addition to its primary science goals, ANTARES was also used to study various phenomena, including cosmic rays \cite{Romanov_2021}, atmospheric neutrinos oscillations \cite{Adri_n_Mart_nez_2012_2}, dark matter \cite{Albert_2022}\cite{Adri_n_Mart_nez_2016}, and the search for exotic particles that emit light, such as magnetic monopoles and nuclearites \cite{20221}\cite{Albert_2023}\cite{spurio2019searches}.
\section{Magnetic Monopoles}
\subsection{Magnetic Monopoles theory}
Particles with a single magnetic pole, known as magnetic monopoles, are suggested by a number of theories that extend particle physics beyond the Standard Model. They are expected to be massive and stable \cite{spurio2019searches},
and are postulated to carry a magnetic charge expressed as a multiple of the Dirac charge, which is given by: \\
\begin{equation}
g_{\mathrm{D}}=\frac{\hbar c}{2 e}=\frac{e}{2 \alpha}=68.5 e
\end{equation}
where $\hbar$ is the Planck constant, $e$ is the elementary electric charge, $c$ is the speed of light in vacuum, and $\alpha$ is the fine structure constant.\\
While Dirac previously demonstrated the compatibility of magnetic monopoles with quantum electrodynamics, 't Hooft \cite{tHooft:1974kcl} and Polyakov \cite{Polyakov:1974ek} demonstrated the indispensability of magnetic monopoles in the context of Grand Unification Theories. This finding led to the conclusion that any unification model, wherein the U(1) subgroup of electromagnetism is integrated into a semi-simple gauge group and is spontaneously broken by the Higgs mechanism, inherently possesses solutions resembling magnetic monopoles. The mass range of magnetic monopoles spans from $10^8$ to $10^{17}$ GeV.

\subsection{Magnetic Monopoles Signal in Neutrino Telescopes}
As a magnetic monopole traverses a medium, it produces light through diverse processes depending on its velocity range, such as direct Cherenkov emission, indirect Cherenkov emission from ejected $\delta$-rays and luminescence. When moving through a medium of phase refractive index $n$, magnetic monopoles generate visible light when their velocity exceeds the Cherenkov threshold $\textit{i.e.}$ $v > c/n \simeq 0.75 c $.\\
The number of Cherenkov photons $N_\gamma$ emitted 
by a magnetic monopole with magnetic charge $ g $, per unit path length $dx$ and unit photon wavelength $d\lambda$ can be described by 
 the Frank and Tamm formula \cite{PhysRev.138.B248}, and is expressed as:

\begin{equation}
\label{eq:tompkins}
\frac{d^2 N_\gamma}{dx d\lambda} = \frac{2\pi\alpha}{\lambda^2}\left(\frac{gn}{e}\right)^2\left(1-\frac{1}{\beta^2n^2}\right)
\end{equation}
where \( \beta = \frac{v}{c} \) is the magnetic monopole’s speed. As results from Eq.\eqref{eq:tompkins}, the quantity of photons emitted by a magnetic monopole is 8200 times greater than that emitted by a particle with an elementary electric charge $ e $ at the same velocity.
Beyond the direct emission of Cherenkov radiation, magnetic monopoles can produce light through an indirect process involving $\delta$-rays that contribute to the total light yield if their velocities reach the Cherenkov threshold. In this case, the number of Cherenkov photons having a wavelength between 300 and 600 nm  per unit path length can be obtained by the following expression \cite{SELTZER1984665}.

\begin{equation}
   \frac{d N_{\gamma}}{d x}=\int_{T_{0}}^{T_{m}} \frac{d^{2} N_{e}}{d T_{e} d x}\left[\int_{T_{0}}^{T_{e}} \frac{d N_{\gamma}}{d x_{e}}\left(\frac{d E_{e}}{d x_{e}}\right)^{-1} d E_{e}\right] d T_{e},
\end{equation}

\noindent where $E_e$ and $N_e$ designate respectively the energy and the number of the $\delta$-rays, 
$dx$ and $dx_e$ are respectively the unit length traveled by a magnetic monopole and an electron,
$T_e$ is the electron kinetic energy varying between its minimum and maximum limits $T_0 = 0.25$ MeV and $T_m$, respectively, as applied in the KYG model \cite{Kazama:1976fm} used in this study: 
\begin{equation}
{ T_{m} = 2\textit{m}_{e}c^{2}\beta^{2}\gamma
^{2},}
\end{equation}

\noindent where ${\gamma}$ represents the Lorentz boost factor.

\begin{figure}[h]
    \centering   \includegraphics[width=0.8\linewidth]{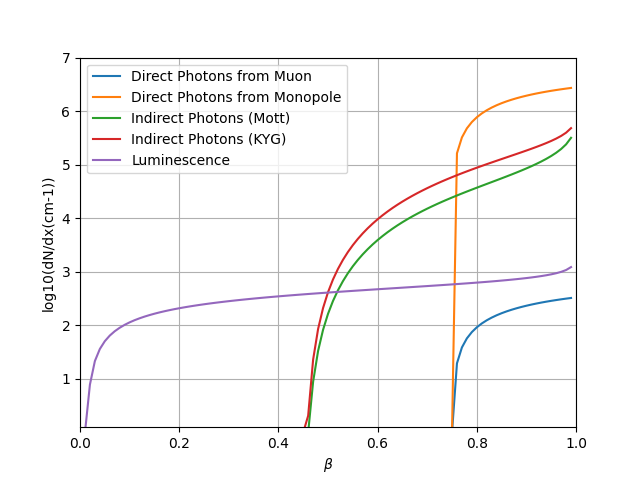}
    \caption{The number of Cherenkov photons emitted per cm in the sea water from a magnetic monopole with Dirac charge (orange line) and from $\delta$-rays
produced along its path for the Mott model (green line) \cite{Ahlen:1976jw}, KYG model (red line) \cite{Kazama:1976fm} and the luminescence process (magenta line) as a function of the velocity of the monopole. The direct Cherenkov emission from a muon is also shown as a comparison reference (blue line).}
\label{fig:yourlabel}
\end{figure}

\noindent In addition to Cherenkov radiation emitted directly and through  secondary $\delta$-rays,  another process observed in neutrino telescopes operating in water is luminescence. This involves energy loss converting into excitation of atomic/molecular levels \cite{refId0}, producing visible light. Luminescence's characteristics, like wavelength spectrum and decay times, vary with the medium's temperature and purity. Unlike the rapid Cherenkov emission, luminescence has slower decay times 
\cite{spurio2019searches}.
The luminescence light yield is determined by multiplying the energy loss of magnetic monopoles (as derived from Refs \cite{refId0} and \cite{Derkaoui:1998uv}) by the luminescence efficiency. It is expected that the luminescence light produced is about one order of magnitude less than that produced by \(\delta\)-rays. The number of photons emitted by a magnetic monopole per unit length in seawater is shown in figure \ref{fig:yourlabel}, following the different mentioned processes as a function of the magnetic monopole's velocity.

\section{Simulation, Trigger and Reconstruction}
This analysis uses Monte Carlo (MC) simulations to generate the magnetic monopoles signal, according to the KYG model. The background includes atmospheric muons and neutrinos. To address fluctuations in environmental conditions within the seawater that affect data acquisition and optical module efficiencies, simulation files containing both signal and background events are produced using the run-by-run simulation strategy \cite{article}. \\
The simulation of magnetic monopoles is divided in 10 equally spaced velocity ranges within the interval of $\beta$ $\in$ $[0.550 - 0.995]$ \cite{ANTONIOLI1997357}. 
For each velocity range, 500 events per run are generated uniformly on the surface of a cylindrical volume around the detector. In accordance with the detector's acceptance criteria \cite{Albert_2021}, direct and $\delta$-ray Cherenkov light emissions are simulated for  photons with a wavelength between 300 and 600 nm covering the sensitive range of the ANTARES photomultipliers \cite{Albert_2018}.\\
The primary background in this study originates from up-going muons generated by atmospheric neutrinos, and down-going atmospheric muons wrongly reconstructed as upgoing. The simulation of atmospheric muons is conducted using the MUPAGE generator \cite{Mupage}. This generator employs parameterizations of the angular and energy distributions of muons in an underwater environment, considering the dependence of muon bundle multiplicity on these distributions \cite{Becherini:2005sr}.

\noindent Local coincidences, which are defined as the occurrence of either two hits on two distinct optical modules of a single storey within 20 ns or one hit of substantial amplitude, usually more than three photo-electrons, constitute the basic selection level of the applied triggers. The trigger used in this research relies on the combination of two local coincidences occurring in adjacent or next to adjacent storeys within 100 or 200 ns, respectively. Only events that meet the trigger suitable for magnetic monopoles are taken into account in this study.\\
The algorithm outlined in \cite{Aguilar_2011} has been slightly adjusted for the event reconstruction. It makes the default hypothesis that particles move at the speed of light. To enhance the sensitivity for magnetic monopoles moving at lower speeds, the technique was updated to allow the track fit to determine the reconstructed particle velocity, $\beta_{\text{reco}}$, as a free parameter. 
The reconstruction algorithm evaluates the quality of the fit using two parameters: \( t\chi^2 \), corresponding to the track fit, and \( b\chi^2 \), associated with the bright-point fit. This study employs two distinct approaches depending on the velocity of the  magnetic monopoles searched for: 
($\beta_\text{reco} = 1$) for fast monopoles exceeding the Cherenkov threshold, and 
($\beta_\text{reco} = free$) for slow monopoles below the threshold.

\section{Data filtering and Signal Selection}
\begin{figure}[htb]
	\centering
	\includegraphics[width=0.75\linewidth]{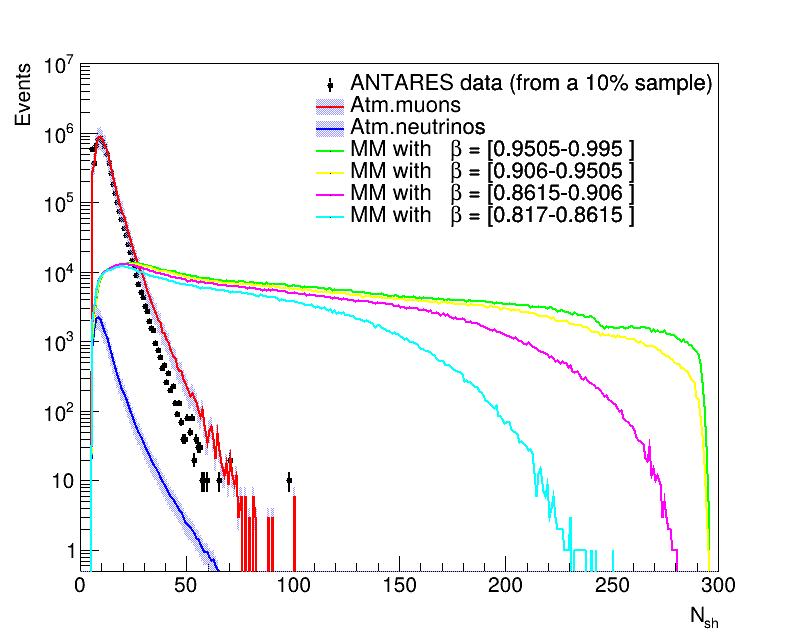}
	\caption{$N_{\text{sh}}$ for atmospheric muons (red histogram), atmospheric neutrinos (blue
		histogram) with an uncertainty band of 35\% (filled in gray) and ANTARES data (from a 10\% sample). For comparison, $N_{\text{sh}}$ distributions for simulated magnetic monopoles are shown for various velocity ranges: $\beta$ $\in$ $[0.9505 - 0.995]$ in green, 
		$\beta$ $\in$ $[0.906 - 0.9505]$ in yellow,
		$\beta$ $\in$ $[0.8615 - 0.906]$ in magenta and
		$\beta$ $\in$ $[0.817 - 0.8615]$ in cyan.}
	
	\label{fig:tchi2nbetaallpr}
\end{figure}

\begin{figure}
	\centering
	\includegraphics[width=0.7\linewidth]{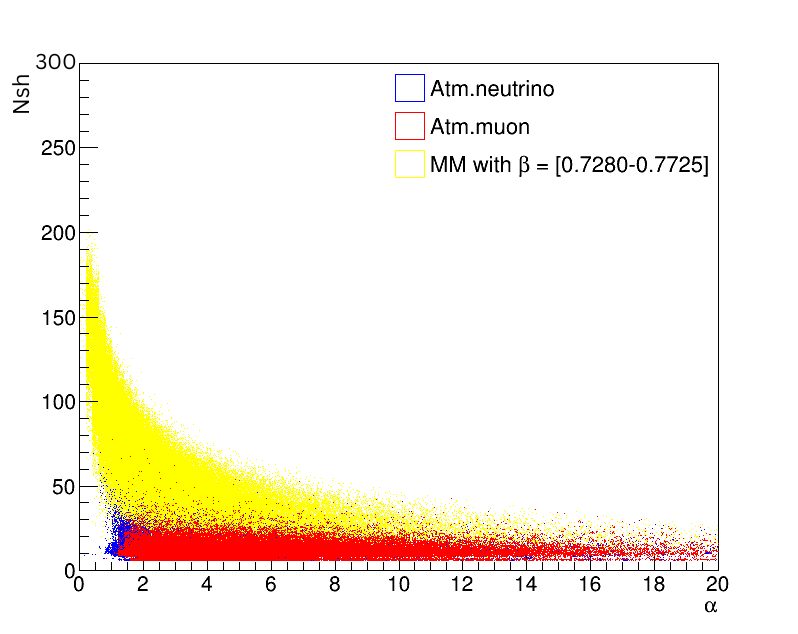}
	\caption{Scatter plot of $\alpha$ versus $N_{\text{sh}}$ for atmospheric muons in red, atmospheric neutrinos in blue and magnetic monopoles with velocity $\beta$ $\in$ $[0.7280-0.7725]$ in yellow.}
	\label{fig:4alphanhit}
\end{figure}

The events of interest in this study are identified by applying a series of selection criteria designed to reduce the background composed of atmospheric muons and neutrinos. The zenith angle of the reconstructed track, denoted as $\theta_{\text{reco}}$, is the primary selection parameter. By using the Earth as a filter, events with $\theta_{\text{reco}} < 90^\circ$ are selected, corresponding to up-going tracks. To enhance the quality of the event reconstruction \cite{Adri_n_Mart_nez_2014}, only reconstructed events involving at least two detector lines are selected. Additionally, a pre-cut of $t\chi^{2}<10$ is applied to retain only well-reconstructed tracks.
\noindent The number of storeys with fired photomultipliers, $N_{\text{sh}}$, is a key parameter for both data and MC reconstruction. Event brightness and reconstruction quality $t \chi^{2}$  are important discriminant variables allowing separation of magnetic monopoles signal from the background. They are combined through the parameter $\alpha$ defined as:
\begin{equation}
\alpha=\frac{t \chi^{2}}{1.3+\left(0.04 \times\left(N_{sh}-N_{d f}\right)\right)^{2}}{,}
\end{equation}
The number of degrees of freedom $N_{df}$ indicates the number of parameters reconstructed and is equal to 5 at high velocities, while it is equal to 6 at low velocities. 
In this analysis, $N_{\text{sh}}$ and $\alpha$ will serve as major discriminant variables for magnetic monopoles event selection in both low and high velocity ranges. The selection of magnetic monopole events against the background is performed using a blinded strategy to prevent bias.
\begin{itemize}
	\item\textbf{High velocity } 
\end{itemize}
The fast magnetic monopole simulation is divided into four intervals and monopoles are generated using $\beta$ in the range [0.8170, 0.9950]. They are then reconstructed with $(\beta_{\text{reco}} = 1)$. To distinguish the signal from the background in this velocity range, relativistic magnetic monopoles are expected to emit significantly more Cherenkov radiation in the detector compared to muons. As illustrated in figure \ref{fig:yourlabel}, the light yield generated by a magnetic monopole is over two orders of magnitude greater than that of a single muon.\\
$N_{\text{sh}}$ and $\alpha$ distributions obtained from the measured data and the MC simulation of atmospheric neutrinos, atmospheric muons and high velocity magnetic monopoles are gathered together (see figures \ref{fig:tchi2nbetaallpr} and \ref{fig:4alphanhit}). This analysis aims to extract the contribution of magnetic monopoles to the measured signals through rejection of muons and neutrinos considered as a background. Optimized  cuts on the main parametrers $\alpha$ and $N_{\text{sh}}$ are used to obtain the best sensitivity to magnetic monopoles.

\begin{itemize}
	\item\textbf{Low velocity} 
\end{itemize}
Slower magnetic monopoles are simulated in the $\beta$ range [0.5500, 0.8170]. This velocity range is used to generate events across six evenly spaced intervals. A free parameter in the reconstruction process, $\beta_{\text{reco}}$, is employed to search for magnetic monopole events that were simulated. Unlike the previous analyses, a relaxation with 5\% was applied on $\beta$ cut in order to enhance the sensitivity of slow magnetic monopole for each velocity. Misreconstructed atmospheric muons are eliminated by implementing additional cuts to the track fit quality parameter. The condition $t\chi^2 \leq b\chi^2$ is specifically employed to preferentially select events that are reconstructed as a track than those that are reconstructed as a bright point \cite{Aguilar_2011}.

\noindent The optimization of the selection parameters $\alpha$ and $N_{\text{sh}}$  is conducted within six bins of $\beta$ in the range $[0.5500-0.8170[$, below the Cherenkov threshold.

\section{Optimization of cuts}

\noindent
As a first step, the expected background is quantified. To deal with the low statistics of atmospheric muons with large $N_{\text{sh}}$, an extrapolation has been made in the signal region by fitting the $N_{\text{sh}}$ histogram of muon with the Landau distribution.
The contribution from this extrapolation, along with the atmospheric
muon rate as defined in Eq.\eqref{eq:atm_muon}, is included in the total background events estimation used for the sensitivity calculation:

\begin{equation}
N_{\text{atm,muon}}(N_{\text{sh,cut}}, \alpha_{\text{cut}}) =
\int_{N_{\text{sh,cut}}}^{\infty} \text{Landau}(N_{\text{sh}}) \, dN_{\text{sh}} \cdot
\frac{\int_{0}^{\alpha_{\text{cut}}} N_{\text{ev}}(\alpha) \, d\alpha}{\int_{0}^{\infty} N_{\text{ev}}(\alpha) \, d\alpha}
\label{eq:atm_muon}
\end{equation}

\noindent Notice that the use of the Landau function extrapolation is a good way to smooth the background expectation that may have large fluctuations due to low statistics. Anyway, its use together with a uniform $\alpha$ distribution could be considered a conservative assumption. To determine the best 90\% C.L. limits for various monopole velocities, the model rejection factor (MRF) \cite{Hill_2003} was used for each monopole velocity and it was optimized separately. The 90\% C.L. sensitivity $S_{90 \%}$ is computed using the Feldman-Cousins \cite{feldman1998unified} formula, taking into account events that result from a Poisonnian distribution:

\begin{equation}
\centering
S_{90 \%}{ [\mathrm{cm}^{-2} \:\mathrm{s}^{-1} \: \mathrm{sr}^{-1}]}
=\frac{\bar{\mu}_{90}\left(n_{b}\right)}{A_{e f f}{\left[\mathrm{cm}^{2}\; \mathrm{sr}\right] }\times T[\mathrm{s}]} ,
\end{equation}

\noindent where \textit{T} is the duration of data taking, { $n_{b}$ representing the number of expected background events in the 90\% C.L. interval} ($\mu_{90}$), $\bar{\mu}_{90}$ and $A_{e f f}$ are defined as:
\begin{equation}
\centering
\bar{\mu}_{90}\left(n_{b}\right)=\sum_{n_{o b s}={0}}^{\infty} \mu_{90} \frac{n_{b}^{n_{o b s}}}{n_{o b s} !} e^{-n_{b}} ,
\end{equation}

\begin{equation}
\centering
A_{e f f}=\frac{n_{M M}}{\Phi_{M M}} ,
\end{equation}

\noindent where $\ n_{M M}$ is the number of magnetic monopoles remaining after cuts, $\Phi_{MM}(\mathrm{cm}^{-2} \mathrm{sr}^{-1})$ represents the flux of generated magnetic monopoles and $n_{obs}$ is the total number of observed events. The MRF method requires changing the cuts until the best sensitivity and minimum flux are found.

\section{Results}
The procedure described above resulted in the optimized cuts shown in table \ref{results_tab} for the different $\beta$ intervals. For the high-velocity intervals, where $\beta_{\text{reco}} = 1$, the expected background was 1.24, or 1.67 for the looser cut, whereas the number of observed events in data was null. For the low-velocity intervals, with $\beta_{\text{reco}}$ left free, the number of expected events ranged from 0.76 for the largest velocity interval up to 1.70 for the lowest one. In these intervals, the number of observed events was null, except for the interval $[0.6490,\,0.6835[$ $\pm\,5\%$, where one event was observed. Overall, the number of observed events was lower than the expected background, which can be understood in terms of the conservative assumptions made for the estimation of the atmospheric muon contribution, as defined in Eq.\eqref{eq:atm_muon}. The final 90\% confidence level (C.L.) upper limits on the magnetic monopole flux in each $\beta$ interval are presented in table \ref{results_tab}. Since no excess over the expected background was observed, the limits were set equal to the sensitivities, following a conservative approach.

\begin{table}[h]
	\centering
	\begin{center}
		\scalebox{0.805}{
			\begin{tabular}{|c|c|c|c|c|c|c|c|}
				\hline
				\multirow{2}{*}{$\beta$ Interval} & \multirow{2}{*}{$\beta_{reco}$} & \multirow{2}{*}{$\beta$ cut } & \multirow{2}{*}{$\alpha$ cut} & \multirow{2}{*}{$\mathrm{N_{sh}} $ cut }    &  \multicolumn{1}{c|}{Flux upper limit}  \\
				 & & & &   &  $\left[\mathrm{cm}^{-2} \; \mathrm{s}^{-1}\; \mathrm{sr}^{-1}\right]$

				\\\hline \hline 
				[0.9505, 0.9950] & 1 & - & $< 0.3$ & $\geq 105$    & $5.6 \times 10^{-19}$  
				\\\hline
				[0.9060, 0.9505[ & 1 & - &$< 0.3$ & $\geq 105 $  &  $6.7 \times 10^{-19}$   
				\\\hline	
				
				[0.8615, 0.9060[ & 1 & - &$< 0.3$ & $\geq 105  $& $1.1 \times 10^{-18}$
				\\\hline
				
				[0.8170, 0.8615[ & 1 & - &$< 0.6$ & $\geq 102 $ &   $2.6 \times 10^{-18}$   
				
				\\\hline \hline				
				[0.7725, 0.8170[ & Fitted & [0.7725, 0.8170[$\pm 5\%$ &$< 1$ & $\geq 124$   & $3.4 \times 10^{-18}$ 
				\\\hline  
				[0.7280, 0.7725[ & Fitted & [0.7280, 0.7725[ $\pm 5\%$  &$< 2.6$ & $\geq 94$ & $3.1 \times 10^{-18}$ 
				\\\hline  
				[0.6835, 0.7280[  & Fitted & [0.6835, 0.7280[ $\pm 5\%$ &$< 4$ & $\geq 78$ & $3.2 \times 10^{-18}$  
				\\\hline 
				[0.6390, 0.6835[ & Fitted & [0.6390, 0.6835[ $\pm 5\%$ &$< 6.6$ & $\geq 66 $& $3.0 \times 10^{-18}$   
				\\\hline 
				[0.5945, 0.6390[ & Fitted & [0.5945, 0.6390[ $\pm 5\%$ &$< 8.2$ & $\geq 54 $&  $5.8 \times 10^{-18}$   
				
				\\\hline 			
				[0.5500, 0.5945[  & Fitted & [0.5500, 0.5945[ $\pm 5\%$ & $< 9$ & $\geq 50 $& $8.0 \times 10^{-18}$

				\\\hline

			\end{tabular}
		}
	\end{center}
	
	\caption{The optimized cuts, the number of expected background events remaining after cuts, the number of observed events, and the 90\% C.L. upper limits on the flux obtained in each $\beta$ range are presented for the 14 years of analyzed ANTARES data, corresponding to 3286 days of livetime.}
	\label{results_tab}
\end{table}

\begin{figure}[htb!]
	\centering
	\includegraphics[width=0.9\linewidth]{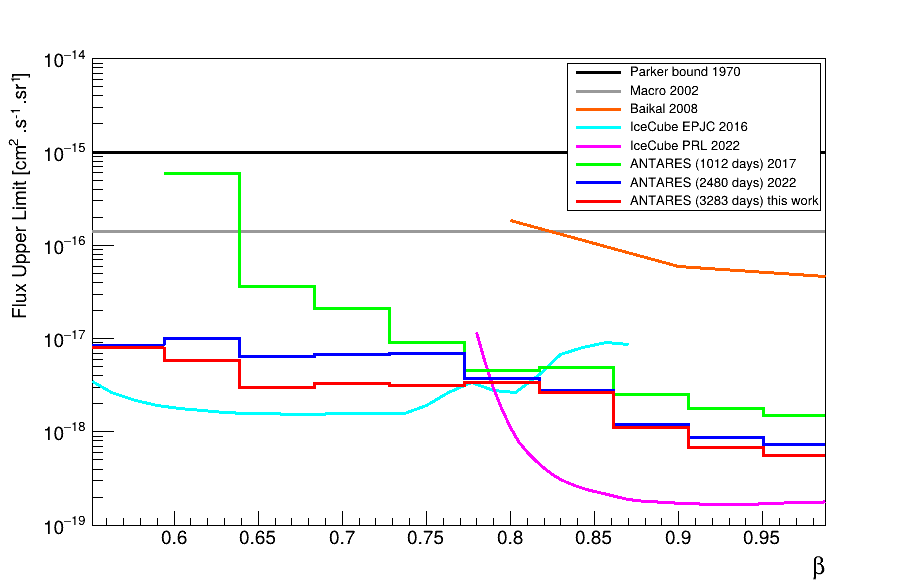}
	\caption{The Flux upper limits obtained with the ANTARES detector corresponding to 14 years of
		analyzed data (3286 days, red line) in comparison with other experiments and results, including: ANTARES
		previous upper limit on the flux (green \cite{Albert_2017} and blue lines \cite{20221}), \ IceCube (cyan  \cite{Aartsen_2016} and magenta lines \cite{Abbasi_2022}), MACRO (gray line \cite{Ambrosio_et_al__2002}) and Baikal (orange line \cite{2008APh....29..366A}), as well as the theoretical Parker bound (black line \cite{osti_4147943}).}
	\label{fig:sensitivity}
\end{figure}

The flux upper limit obtained in this analysis is shown in figure \ref{fig:sensitivity} along with results from previous ANTARES analyses \cite{Albert_2017}\cite{20221} and other experiments: (IceCube \cite{Abbasi_2022}, MACRO \cite{Ambrosio_et_al__2002} and Baikal \cite{2008APh....29..366A}).
\newpage
\noindent After applying the final cuts to the unblinded data, the observed event in the third bin of $\beta$ is characterized by $\mathrm{N_{sh}}$ = 66, $\alpha$ = 3.83 and a zenith angle of $88^{\circ}$. This result is consistent with the expected background. The event was detected near the bottom of an outer line, in close proximity to the detector and due to the different velocity used in the fit for slow monopoles, the event is reconstructed as upward-going in this analysis. Studying this event in detail, is most probably a down-going event (muon bundle), based on its physical properties.

\section{Conclusion}

A search for relativistic magnetic monopoles  has been conducted using the ANTARES neutrino telescope, utilizing a substantial dataset spanning 14 years (2008–2022). The analysis employed an optimized simulation strategy based on the KYG model, and no evidence of magnetic monopoles was found. However, new upper limits on the magnetic monopoles flux have been obtained, surpassing previous ANTARES results, particularly for lower velocities. Future experiments such as KM3NeT \cite{AdrianMartinez2016_1}, with larger detector volumes and improved detection technologies, are expected to achieve even greater sensitivity to these putative particles.

\newpage
\section*{Acknowledgements}
The authors acknowledge the financial support of the funding agencies:
Centre National de la Recherche Scientifique (CNRS), Commissariat \`a
l'\'ener\-gie atomique et aux \'energies alternatives (CEA),
Commission Europ\'eenne (FEDER fund and Marie Curie Program),
LabEx UnivEarthS (ANR-10-LABX-0023 and ANR-18-IDEX-0001),
R\'egion Alsace (contrat CPER), R\'egion Provence-Alpes-C\^ote d'Azur,
D\'e\-par\-tement du Var and Ville de La
Seyne-sur-Mer, France;
Bundesministerium f\"ur Bildung und Forschung
(BMBF), Germany; 
Istituto Nazionale di Fisica Nucleare (INFN), Italy;
Nederlandse organisatie voor Wetenschappelijk Onderzoek (NWO), the Netherlands;
Romanian Ministry of Research, Innovation and Digitalisation (MCID), Romania;
MCIN for PID2021-124591NB-C41, -C42, -C43 and PDC2023-145913-I00 funded by MCIN/AEI/10.13039/501100011033 and by “ERDF A way of making Europe”, for ASFAE/2022/014 and ASFAE/2022 /023 with funding from the EU NextGenerationEU (PRTR-C17.I01) and Generalitat Valenciana, for Grant AST22\_6.2 with funding from Consejer\'{\i}a de Universidad, Investigaci\'on e Innovaci\'on and Gobierno de Espa\~na and European Union - NextGenerationEU, for CSIC-INFRA23013 and for CNS2023-144099, Generalitat Valenciana for CIDEGENT/2020/049, CIDEGENT/2021/23, CIDEIG/2023/20, ESGENT2024/24, CIPROM/2023/51, GRISOLIAP/2021/192 and INNVA1/2024/110 (IVACE+i), Spain;
Ministry of Higher Education, Scientific Research and Innovation, Morocco, and the Arab Fund for Economic and Social Development, Kuwait.
We also acknowledge the technical support of Ifremer, AIM and Foselev Marine
for the sea operation and the CC-IN2P3 for the computing facilities.

 \bibliographystyle{elsarticle-num} 
 \bibliography{biblio.bib}





\end{document}